# Free Fall of a Cylindrical Mass Electrically Charged Bodies

C. J. de Matos[*]

*ESA-ESTEC, Directorate of Scientific Programmes, PO Box 299, NL-2200 AG Noordwijk, The Netherlands*

M. Tajmar[†]

*Austrian Research Centers Seibersdorf, A-2444 Seibersdorf, Austria*

**Abstract**

The gravitational Poynting vector, $\vec{S}_g = \frac{c^2}{4\pi G}\vec{\gamma}\times\vec{B}_g$, provides a mechanism for the transfer of gravitational energy to a system of falling objects. In the following we will show that the gravitational poynting vector together with the electromagnetic Poynting vector provides a mechanism to explain how massive electrically charged bodies acquire kinetic energy during a free fall. We will demontrate that falling electrically charged masses violate the Galilean law of universal free fall. An experiment is suggested to confirm or not the predicted phenomena.

[*] Staff Member, Science Management Communication Division, Phone: +31-71-565 3460, Fax: +31-71-565 4101, E-mail: clovis.de.matos@esa.int
[†] Research Scientist, Space Propulsion, Phone: +43-50550-3142, Fax: +43-50550-3366, E-mail: martin.tajmar@arcs.ac.at

## Free fall of an electrically neutral massive cylinder

Consider a cylinder of length $\ell$, radius $a$, mass $\rho$ per unit volume, moving downward with the influence of a gravitational field $\vec{\gamma}$.

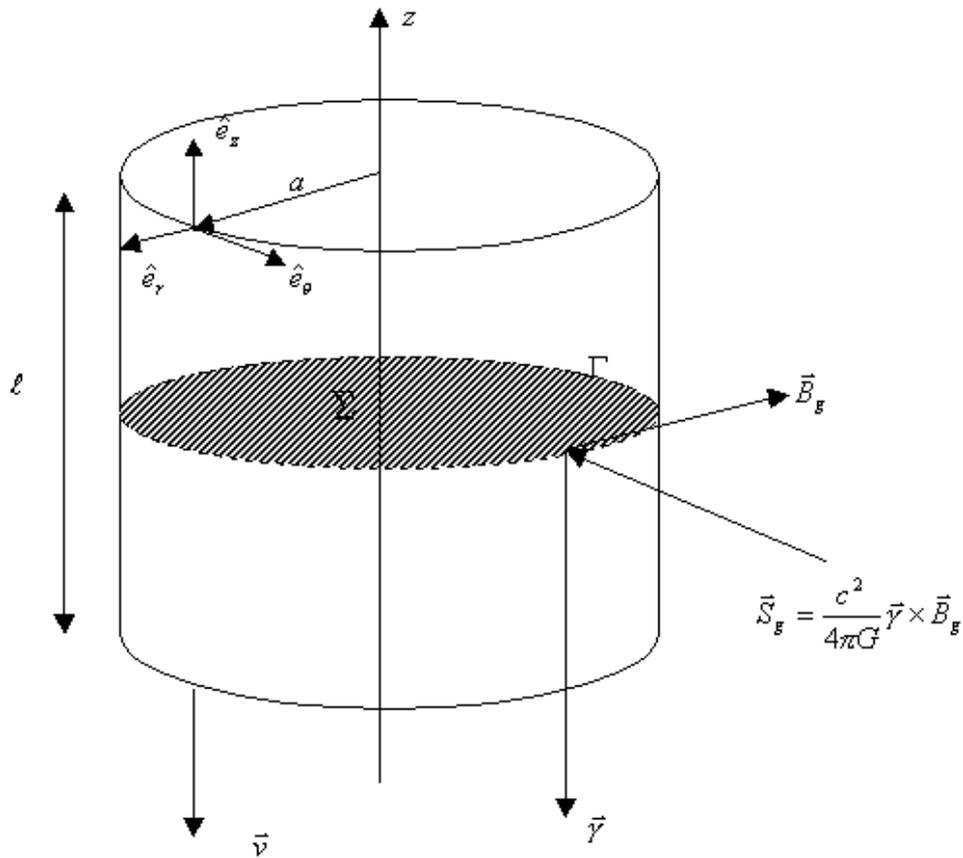

**Figure 1 Non-rotating cylinder falling in a gravitational field**

Let us use cylindrical coordinates for describing the motion of the cyclinder and the corresponding energy relations. The gravitational field around the cylinder is equal to the acceleration of Earth gravity $\vec{\gamma}$ and is directed downwards:

$$\vec{\gamma} = -\gamma \hat{e}_z \qquad (1)$$

Let the velocity of the cylinder be $\vec{v}$. The cylinder constitutes a mass current which is given by:

$$\vec{j} = \rho \vec{v} = -\frac{m\vec{v}}{\pi a^2 \ell} \hat{e}_z. \quad (2)$$

The gravitomagnetic field at the surface of the cylinder will be:

$$Rot\, \vec{B}_g = -\frac{4\pi G}{c^2} \vec{j}$$

$$\oiint_\Sigma Rot\, \vec{B}_g \cdot d\vec{\sigma} = -\frac{4\pi G}{c^2} \oiint_\Sigma \vec{j} \cdot d\vec{\sigma}$$

$$\oint_\Gamma \vec{B}_g \cdot d\vec{s} = -\frac{4\pi G}{c^2} \oiint_\Sigma \vec{j} \cdot d\vec{\sigma} \quad (3)$$

$$2\pi a B_g = -\frac{4\pi G}{c^2} j \pi a^2 \quad (4)$$

Inserting equation (2) into equation (4)

$$B_g = -\frac{2G}{c^2} \frac{mv}{\ell a} \quad (5)$$

The minus sign appearing in equation (5) is due to the fact that the direction of $B_g$ is given by the "left hand rule" since, contrary to electric case, like masses attract, therefore in the given situation represented in figure 1

$$\vec{B}_g = \frac{2G}{c^2} \frac{mv}{\ell a} \hat{e}_\theta \quad (6)$$

The gravitational Poynting vector [2], [3] at the cylinder surface is:

$$\vec{S}_g = \frac{c^2}{4\pi G} \vec{\gamma} \times \vec{B}_g = \frac{mv\gamma}{2\pi a \ell} \hat{n}_{in} \quad (7)$$

where $\hat{n}_{in} = -\hat{e}_r$ is a unit vector normal to the cylinder's surface and directed into the cylinder. Multiplying $\vec{S}_g$ by the area of the cylinder, we obain for the rate of gravitational energy influx into the cylinder:

$$\frac{dU}{dt} = \frac{mv\gamma}{2\pi a \ell} 2\pi a \ell = mv\gamma \quad (8)$$

The rate at which the kinetic energy of the cylinder increases is:

$$\frac{dU}{dt} = \frac{d}{dt}\left(\frac{mv^2}{2}\right) = mv\frac{dv}{dt} = mv\dot{v} \qquad (9)$$

We just have shown that the rate at which the kinetic energy of the cylinder increases is completely accounted by the influx of gravitational energy into the cylinder [1].

**Free fall of an electrically charged massive cylinder**

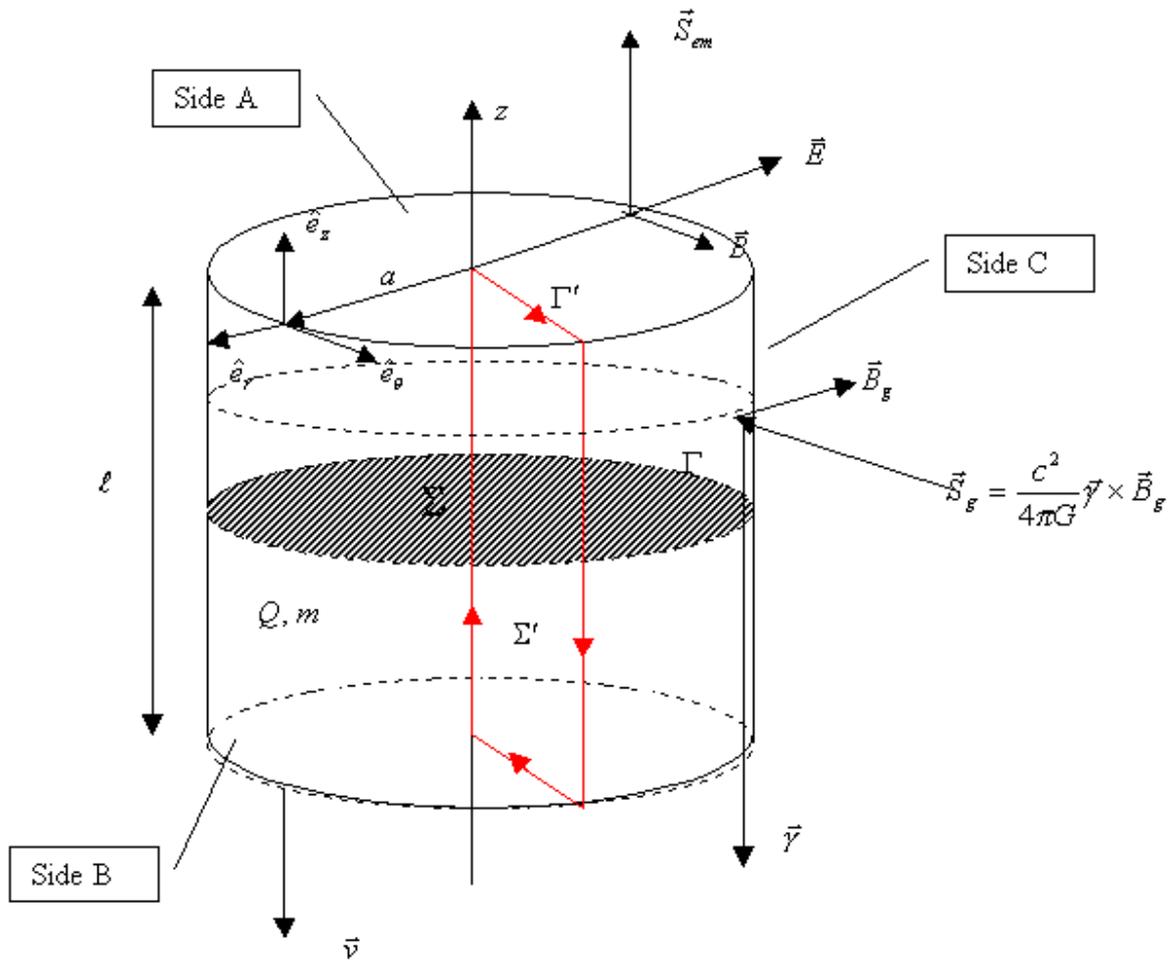

**Figure 2 Free fall of cylinder with electric charge Q and mass "m", up side-A, down side-B, cylindrical side-C**

The cylinder carries an electric charge $Q$. During its fall the cylinder will generate a magnetic field due to the electric charge current associated with its motion. The electric density current vector will be given by:

$$\vec{j} = \rho\,\vec{v}$$

$$\vec{j} = \frac{Q}{\pi a^2 \ell}\vec{v}$$

$$\vec{j}_\Omega = -\frac{Q}{\pi a^2 \ell}v\hat{e}_z \quad (10)$$

The magnetic field at the surface of the cylinder due to the electric charge current generated by the fall of the cylinder is:

$$\oint_\Gamma \vec{B}\cdot d\vec{s} = \mu_0 \iint_\Sigma \vec{j}\cdot d\vec{\Sigma} \quad (11)$$

Doing (10) into (11)

$$B 2\pi a = \mu_0 \frac{Qv}{\pi a^2 \ell}\pi a^2$$

$$\vec{B} = -\frac{\mu_0}{2\pi}\frac{Qv}{a\ell}\hat{e}_\theta \quad (12)$$

The electric field generated by the variation in time of the flux of $\vec{B}$ through section $\Sigma'$ generates by induction an electric field along the contour $\Gamma'$:

$$\oint_{\Gamma'} \vec{E}\cdot d\vec{l} = -\frac{d}{dt}\iint_{\Sigma'} \vec{B}\cdot d\vec{\sigma}$$

$$E(2a+2\ell) = -\frac{dB}{dt}a\ell \quad (13)$$

Inserting (12) into (13) we end with:

$$E = -\frac{\mu_0}{4\pi}\frac{1}{a+\ell}Q\dot{v} \quad (14)$$

From (12) and (14) we compute the electromagnetic Poynting vector for side A, B and C

$$\vec{S}_{em} = \frac{1}{\mu_0}\vec{B}\times\vec{E} = \frac{\mu_0}{8\pi^2}\frac{Q^2 v\dot{v}}{a\ell(a+\ell)}\hat{n}_{out} \quad (15)$$

$\hat{n}_{out}$ is a unit vector normal to sides A, B, and C and directed out-ward the cylinder. This is the power per unit surface radiated while the cylinder is accelerating. The principle of equivalence states that if the cylinder is at rest with respect to a reference frame which is uniformly accelerating upwards (with respect to the laboratory) with acceleration $\dot{\vec{v}} = \gamma \hat{e}_z$, the cylinder will radiate (with respect to the laboratory) according to the following Poynting vector:

$$\vec{S}_{em} = \frac{1}{\mu_0} \vec{B} \times \vec{E} = \frac{\mu_0}{8\pi^2} \frac{Q^2 v \gamma}{a\ell(a+\ell)} \hat{n}_{out}. \quad (16)$$

Therefore to comply with the principle of equivalence, we shall take in equation (15), $\dot{v} = \gamma$ [5]. From (16) we deduce that the rate at which electromagnetic energy is radiated through sides A, B, and C is given by:

$$\left.\frac{dU}{dt}\right|_{em} = S_{em}(2\pi a^2 + 2\pi a\ell) \quad (17)$$

doing (16) into (17)

$$\left.\frac{dU}{dt}\right|_{em} = -\frac{\mu_0}{4\pi} \frac{Q^2 v \gamma}{\ell} \quad (18)$$

The rate at which the gravitational energy of the cylinder increases is given from equation (8) above

$$\left.\frac{dU}{dt}\right|_{\gamma} = \frac{mv\gamma}{2\pi a\ell} 2\pi a\ell = mv\gamma \quad (19)$$

The rate at which the kinetic energy of the cylinder increases is:

$$\left.\frac{dU}{dt}\right|_{T} = \frac{d}{dt}\left(\frac{mv^2}{2}\right) = mv\frac{dv}{dt} = mv\dot{v} \quad (20)$$

The law of conservation of energy implies that:

$$\left.\frac{dU}{dt}\right|_{em} + \left.\frac{dU}{dt}\right|_{\gamma} = \left.\frac{dU}{dt}\right|_{T} \quad (21)$$

Doing (18), (19) and (20) into (21) we get:

$$\dot{v} = \gamma\left(1 - \frac{\mu_0}{4\pi} \frac{Q^2}{m\ell}\right) \quad (22)$$

Equation (28) shows that the acceleration at which an electrically charged cylinder will fall depends on its electric charge, on its mass and on its length. For $Q = 0$ we have the usual result

$\dot{v} = \gamma$. For $m = 10\,[gr]$, $\ell = 10\,[cm]$, $Q = 100\,[C]$ we will have $\dot{v} = 0$. For these values the cylinder would not be able to fall! However to avoid disruption currents for such an high value of electric charge is a technological challenge.

Equation (27) is very similar to the result of Petkov [4] for the free fall of an electric dipole (oriented anti-parallel to the earth gravitational field):

$$\dot{v} = \gamma\left(1 - \frac{\mu_0}{4\pi}\frac{Q^2}{md}\right) \quad (23)$$

Where $d$ is the separation distance between the two electric charges $|Q|$ forming the dipole, $\gamma$ is the earth gravitational acceleration and $m$ is the total mass of the dipole. For an electric dipole perpendicular to the earth gravitational field Petkov gets:

$$\dot{v} = \gamma\left(1 - \frac{\mu_0}{2\pi}\frac{Q^2}{md}\right) \quad (24)$$

Therefore, based on the equivalence between equation (22) and (23), we deduce that the free fall of a cylindrical mass electrically charged is similar to the free fall of an electric dipole oriented anti-parallel to the earth gravitational field.

# Conclusion

Equations (22), (23) and (24) violate the universal law of Galilean free fall which states that the acceleration of a massive body falling in the Earth gravitational field does not depend on its mass. To test equation (22) and (24) we propose to measure the time of fall of charged cylindrical capacitors (see Figure 3), and compare it with the time of fall of similar uncharged capacitors.

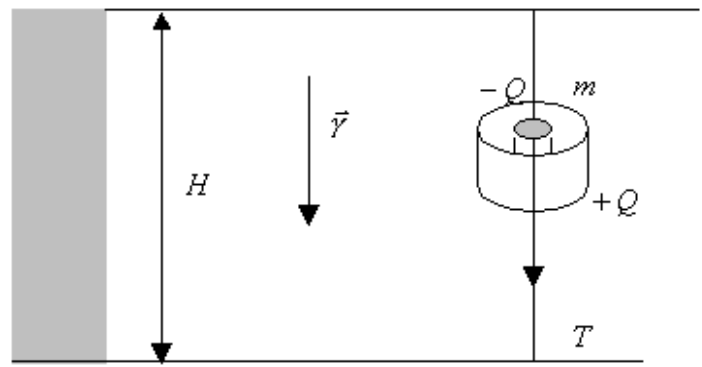

**Figure 3  Falling Capacitor**

Following the rational presented in the section above, the phenomenon described by equation (22) should happen either in a reference frame at rest in an external gravitational field or inside a uniformly accelerated reference frame, therefore we are not able to use this phenomenon to distinguish between both situations. Consequently equation (22) do not violate the principle of equivalence.

**References**

[1] P. Krumm, D. Bedford, "The Gravitational Poynting Vector and Energy Transfer", Am. J. Phys, 55 (4), April 1987 pp 362-363

[2] L. M. de Menezes, "Gravitational Analog of the Electromagnetic Poynting Vector", 29th Jan 1998, gr-qc/9801095

[3] C. de Matos, M. Tajmar, "Gravitational Poynting Vector and Gravitational Larmor Theorem in Rotating Bodies with Angular Acceleration", submitted to Los Alamos Archive.

[4] V. Petkov, "Propulsion through Electromagnetic self-sustained Acceleration", AIAA-99-2144, gr-qc/9906059

[5] N. Soker, " Radiation from an Accelerated Charge and the Principle of Equivalence", "NASA Breakthrough Propulsion Physics Workshop Proceedings", (January 1999), NASA / CP - 1999-208694, pp 427-440